\documentclass[conference]{IEEEtran}

\ifCLASSINFOpdf
\else
\fi
\usepackage[english]{babel}
\usepackage[T1]{fontenc}

\usepackage[utf8x]{inputenc}
\usepackage{cite}
\usepackage{graphicx}
\usepackage{color}
\usepackage{amssymb}
\usepackage{amsthm}
\usepackage{amsxtra}
\usepackage{amsmath}
\usepackage{multirow}
\usepackage{epsfig}
\usepackage{comment}
\usepackage{subfigure}
\usepackage{psfrag}
\usepackage{url}
\usepackage{textcomp}
\usepackage{enumerate}
\usepackage[printonlyused]{acronym}
\usepackage{epstopdf}
\usepackage{balance}
\usepackage{algorithm}
\usepackage{algpseudocode}
\usepackage{nomencl}
\usepackage{units}
\usepackage{pstricks, pst-plot, pst-grad, pstricks-add, pst-node, pstricks-add}
\algnewcommand{\Inputs}[1]{%
  \State \textbf{Inputs:}
   \hspace*{\algorithmicindent}\parbox[t]{.8\linewidth}{\raggedright #1}
}
\algnewcommand{\Initialize}[1]{%
  \State \textbf{Initialization:}
  \Statex \hspace*{\algorithmicindent}\parbox[t]{.8\linewidth}{\raggedright #1}
}
\algnewcommand{\Output}[1]{%
  \State \textbf{Output:}
   \hspace*{\algorithmicindent}\parbox[t]{.8\linewidth}{\raggedright #1}
}
\algnewcommand{\Update}[1]{%
  \State \textbf{Update:}
   \hspace*{\algorithmicindent}\parbox[t]{.8\linewidth}{\raggedright #1}
}
\usepackage{pifont}
\usepackage{glossaries}
\usepackage{siunitx}

\newrgbcolor{darkgreen}{0.2 0.6 0.2}
\newrgbcolor{darkred}{0.6 0.2 0.2}
\newrgbcolor{darkblue}{0.2 0.2 0.6}
\newrgbcolor{purple}{0.2 0.1 0.9}
\begin{document}
%

\title{Impact of Short Blocklength Coding on Stability of Control Systems in Industry 4.0}

\author{
\IEEEauthorblockN{Shreya Tayade\IEEEauthorrefmark{1},
Peter Rost\IEEEauthorrefmark{2},
Andreas Maeder\IEEEauthorrefmark{2} and 
Hans D. Schotten\IEEEauthorrefmark{1}}
\IEEEauthorblockA{\IEEEauthorrefmark{1}Intelligent Networks Research Group, German Research Center for Artificial Intelligence, Kaiserslautern, Germany\\
Email: \{Shreya.Tayade, Hans\_Dieter.Schotten\} @dfki.de}
\IEEEauthorblockA{\IEEEauthorrefmark{2}Nokia Bell Labs,
Munich, Germany\\
Email: \{peter.m.rost, andreas.maeder\}@nokia-bell-labs.com}
}

\maketitle
\makeatletter
\def\ps@IEEEtitlepagestyle{
  \def\@oddfoot{\mycopyrightnotice}
  \def\@evenfoot{}
}
 \def\mycopyrightnotice{
   {\footnotesize
   \begin{minipage}{\textwidth}
   \centering
  Copyright~\copyright~ IEEE ICC 2020. Personal use of this material is permitted. Permission from IEEE must be obtained for all other uses, in any current or future media, including reprinting/republishing this material for advertising or promotional purposes, creating new collective works, for resale or redistribution to servers or lists, or reuse of any copyrighted component of this work in other works. This is not the final version of paper. 
   \end{minipage}
   }
 }
 


\begin{abstract}
With the advent of 5G and beyond, using wireless communication for closed-loop control and automation processes is one of the main aspects of the envisioned Industry 4.0. In this regard, a major challenge is to ensure robustness and stability of control system over unreliable wireless channels. One of the main uses cases in this context is Guided Vehicle (GV) control in a future factory. Specifically, we consider a system where the GV controller is placed in an edge cloud in the factory network infrastructure. In industrial control, short packets are exchanged between the controller and the actuator. Therefore, in this case, Shannon's assumption for an infinite block length is not applicable. Considering a Finite Block Length (FBL), we analyse the impact of short blocklength coding and control parameters such as sampling time and the GV's velocity on stability of the control system. The coding rate required to achieve a stable GV system when controlled over a time-correlated Rayleigh fading channel is determined. The results illustrate that adapting the control parameters can lower the strict communication requirements. It reveals that a constant stability performance can be achieved even at higher coding rate by increasing the GV's velocity. Moreover, this paper also provides the maximum number of control systems that can be served seamlessly over the available communication resources.  
\end{abstract}


\IEEEpeerreviewmaketitle

\section{Introduction} 
Many new use-cases have emerged with the launch of 5G technology. 5G provides an ultra-low latency and high reliability communication, thus accelerating the usage of wireless communication in industrial automation. 
Currently, in industrial automation the control processes communicate over a wired Ethernet based Time Sensitive Network (TSN). The use-cases like remote machine control, tele-operation and monitoring of a factory etc. would require a centralized control and communication to multiple devices over a wireless network. Therefore, the wireless communication systems need to be enhance and adapted as per the requirements of the industrial applications. 

The performance of a Wireless Network Control System (WNCS), is highly sensitive to the wireless channel conditions. Unlike fixed-line communication, the time varying wireless channel induces several uncertainties such as shadowing, path-loss, fading etc. that might cause packet loss and a delay in the network. The uncertainty in a wireless channel causes problems, especially in a closed loop feedback systems, considering the strict latency and reliability requirements. To fulfill the requirements of industrial control, a communication system needs to be designed considering the control system aspects. 

The cross-layer design of communication and control is well known and studied rigorously in the past decade \cite{inbook, 4118476foundation,1310480NoisyChannel,1661825anytime,1333206stochasticcontrol}. The impact of the communication network on the stability of control systems is shown in \cite{inbook, AGVSCC}. The design of an optimal encoder, decoder and controllers to maintain stable control is provided in \cite{1310461comm_constraints,1310480NoisyChannel,1661825anytime,1333206stochasticcontrol}. In \cite{1310480NoisyChannel} the stability criterion is evaluated as per the distortion generated due to the source encoder and decoder. The author in \cite{1310461comm_constraints} evaluates the minimum data rate required to maintain stability. The majority of research work in \cite{4118476foundation,1310480NoisyChannel,1661825anytime,1333206stochasticcontrol, 1310461comm_constraints} evaluate the performance of a simple generic LTI control system. However, in practice the control systems are non-linear, time-varying and inhomogeneous. Also, the stability is evaluated considering the channel models like binary erasure channel, delayed communication channel and memoryless Gaussian channel. In this paper, we consider a Rayleigh fading channel for the analysis. 

In \cite{AGVSCC, anytime} the stability criterion over a Rayleigh fading channel is evaluated. The inter-dependency of the non-linear control and communication system is illustrated in \cite{AGVSCC} by considering a control system of an Autonomous Guided Vehicle (AGV) in a factory. It reveals that the control system performance degrades, if the consecutive control packets are lost, assuming an infinite block length and Shannon's capacity.    
Also, the authors in \cite{anytime} evaluate the minimum required SNR to retain the stability of a Linear Time Invariant (LTI) control system over a Rayleigh fading channel. The analysis considers a finite block length for long blocklength LDPC codes. However, in industrial communication, short data packets are transmitted from sensor to the controller and back to the actuator. Most of the research and analysis in networked controlled system is based on Shannon's capacity for an infinite block length, which is not accurate, considering the short packets in an industrial control. Therefore, we consider a finite block length regime for short block codes to evaluate the performance of control system. 

Moreover, for an industrial application, for example motion control, short data packets must be transmitted frequently and with low packet error rate. Packet loss for longer duration may drive the feedback-control system to instability. Recently, in \cite{Ayan:2019:AVV:3302509.3311050} the stability criterion is determined considering the new metrics as Age of Information (AoI) and Value of Information (VoI). AoI determines the freshness of the control information, while, VoI reveals the reduction of uncertainty in the information. The control system is more likely to become unstable as the AoI and the packet loss is higher. Therefore it is necessary to make a reliable packet transmission prior to the AoI threshold is reached. The reliability of the transmission can be increased by lowering the coding rate. However, transmitting the data with lower coding rate unnecessarily, leads to an inefficient use of the available resources. Resource efficiency becomes crucial if multiple users are scheduled from the same resource pool. The authors in \cite{mikhail, Ayan:2019:AVV:3302509.3311050} address the resource allocation problem considering the control parameters for a simple LTI system. 
 
\begin{figure*}[h!]
    \centering
    \caption{System model: Wireless control systems}
    \label{fig:ch3:1}
\end{figure*}
 
 The objective of this paper is to study the impact of short block codes and coding rate, on the performance of control system over a time correlated Rayleigh fading channel. 
 The required coding rate for a stable non-linear control system  is determined, ensuring efficient usage of resources. In Section \ref{sec: system:model}, we describe the wireless system and the control system of a GV. The stability performance metric  \textit{Probability of instability} is evaluated in section \ref{subsec:instability}. Moreover, we evaluate the admissible coding rate and the maximum number of admissible control systems in Section~\ref{sec:ch5:results}. Finally, the conclusions are discussed in Section~\ref{sec:ch5:conclusion}.

 
\section{System Model} \label{sec: system:model} 
We assume an edge cloud controlling $N$ GVs in a factory. The control system of the GV is a discrete, non-linear, time-varying and inhomogeneous system. The GVs are controlled over a wireless channel as shown in Fig.~\ref{fig:ch3:1}. The control updates are sent to the actuators in downlink and the GV's position update is fed back to the controllers over an uplink channel. 

\subsection{Control System}
 A GV has to trace the planned reference track in time $T$. The reference track is defined for each time step $k$ as $X_r(k)$ = $[x_r(k); y_r(k); \theta_r(k) ]$, where $x_r$ and $y_r$ represents the spatial coordinates and $\theta_r$ is the orientation of the GV. The scheduler allocates the resources equally to the $N$ GVs from the shared resource pool. 
 The control input $u(k) = [\nu(k) ; \omega(k)]$ sent from the controller to the $i^\text{th}$ GV, consists of an intended translational velocity $\nu$ and rotational velocity $\omega$ at each time instant $k$. The control inputs are applied to an actuator at the interval of time $T_s$, which is the sampling time of the control system. A new position $X_c(k+1)$ attained by a GV after applying the control inputs is fed back to the controller and the error is evaluated as given in \cite{AGVSCC}. The position update $X_c(k+1)$ consists of an $x$ and $y$ coordinate and the current orientation of vehicle $\theta_c$. 
The position of a GV at $k+1$ time step after applying the control inputs $u(k)$ is 
\begin{align}
X_c(k+1) & =  X_c(k) +  T_s \cdot \mathbf{J}(k) \cdot u(k),
\label{eq:ch3.ieee.20}
\end{align}
where $J(k)$ is given as
\begin{align}\label{eq:ch3.ieee.19}
\mathbf{J}(k) = \left(\begin{array}{cc} \cos\theta_c(k)& 0\\ \sin\theta_c(k) & 0 \\ 0 & 1\end{array}\right). \end{align}

The control inputs for $k+1$ time instant are generated based on the error $\epsilon(k)$ of a GV. The error as the difference between the actual and reference position $X_c(k)$ and $X_r(k)$, resp., is evaluated as in \cite{AGVSCC}:
\begin{eqnarray}
        \epsilon(k)
    &=& \left(\begin{array}{ccc}\cos\theta_c(k) & \sin\theta_c(k) & 0\\-\sin\theta_c(k) & \cos\theta_c(k) & 0\\0 & 0 & 1\end{array}\right) \left(X_r(k) - X_c(k)\right) \nonumber \\
     &=& \mathbf{T_e}(k) \left(X_r(k) - X_c(k)\right),
   \label{eq:ch3.ieee.10}  
\end{eqnarray}
where $\mathbf{T_e(k)}$ is the rotational matrix, and $\epsilon(k) = \left[x_e(k) ;y_e(k); \theta_e(k) \right]$, where $x_e(k)$, $y_e(k)$ is the error determined as a result of difference in x and y coordinates resp. and $\theta_e(k)$ is the error in the orientation of a GV.

\paragraph*{Stability of feedback control system}
In a cloud based wireless control feedback system, the stability depends on the uplink and downlink channel conditions. Channel outages in the uplink and downlink communication may lead to an unstable control system.  For the sake of brevity, we assume that the uplink channel is perfect and hence does not cause any outages. The stability criterion for a cloud based GV control for downlink channel outages is derived in \cite[eq(15)]{AGVSCC}. A control system is stable if the consecutive control packet loss does not exceed the outage tolerance $n_{max}$. Outage tolerance is the threshold that determines the maximum number of consecutive packet loss that a control system can sustain.

The outage tolerance of a control system depends on the control system sampling time $T_s$ and GV's velocity $\nu$. In case of higher $T_s$, the outage tolerance decreases. The increase in sampling time implies higher rate of change between the two consecutive control updates. Therefore, the system can tolerate lower number of consecutive control packet loss. The outage tolerance of GV for a given sampling time is shown in \cite[Fig.3]{AGVSCC}.

\subsection{Wireless system}
The control information is sent to an actuator over a wireless channel with $D_i$ data bits in the downlink. The $D_i$ data bits are encoded with coding rate $R = \frac{D_i}{L}$. The block channel encoder encodes $D_i$ data bits to $L$ encoded bits. In the industrial communication the packets are short, approximately 100-500 bits, therefore we assume that a single block of length $L$ bits is transmitted per $T_s$ in a control packet. The encoded bits are modulated as per the chosen modulation scheme and the symbols are sent over a time correlated fading channel. The control packets transmitted to the actuators might get lost due to downlink channel outages.
We assume that all the symbols in a single block experiences the same channel gain, and the channel gains over the adjacent blocks are correlated.  An OFDMA system for downlink communication is considered. The least time interval at which control updates can be transmitted to an actuator is every Transmit Time Interval (TTI) , i.e.\, \unit[1]{ms}. $T_s$ of the control system is assumed to be the multiple of TTI. The control updates must be received and decoded successfully by the actuators within time $T_s$. If the updates are not received in time $T_s$, the outdated control information is applied to an actuator. 

\section{Probability of instability over time correlated channel}\label{subsec:instability} 
The performance of a wireless network can be evaluated based on the reliability, end-end latency, and availability. However, the same performance metrics cannot completely describe the WNCS. The operation of a control system is determined by its sampling time (discrete control system), stability, delay and the control information. Therefore, considering both the wireless and the control system aspects, a metric \lq Probability of instability\rq is defined to determine the robustness of a WNCS. It is the likelihood that a WNCS becomes unstable when controlled over a time correlated fading channel. The GV control system becomes unstable if $n$ consecutive control packets are lost, given $n > n_{max}$. The probability that a control packet fails depends on several factors such as the SNR, coding rate, blocklength etc. 
The bounds on the block error probability in a FBL regime is given by Gallager's error exponent and random coding union bound \cite{FBL}. According to the simulation performed for short blocklength with real codes, a saddle point approximation of random coding union bound \cite{Asymptotics} shows the closest performance. The error probability $P_e$ for a block of length $L$ is 

\begin{align}
    P_e(1) = P_o(R) + \frac{\log(L)}{L} \phi \log(R) + \frac{1}{L} \phi_0(R), \\
    \text{where} \quad P_o(R) = 1 - \exp(\frac{-e^R -1}{\gamma}), \\
    \phi = \frac{-e^R}{2\gamma} \exp(\frac{-e^R -1}{\gamma}),  \\
    \phi_0 = \frac{e^R}{\gamma} \exp(\frac{-e^R -1}{\gamma}) \nonumber \\ 
    \left( 2 - \frac{-e^R -1}{\gamma}  + \log\left[\frac{1}{\sqrt{2\pi e(1-e^{-2R})}}\right]  \right),
\end{align}
and $\gamma$ is the instantaneous SNR per block \cite{Asymptotics}. $P_e(1)$ is the probability that single packet transmission fails.  

The average error probability is   
\begin{align}
     \mathbb{E}_\gamma \lbrace P_e(1) \rbrace
\end{align}
the $\mathbb{E}_\gamma$ is expectation over $\gamma$ which is Rayleigh distributed. 

Since the control packets are transmitted frequently i.\,e. at shorter $T_s$, the channel does not vary significantly for next packet transmission. Therefore, the conditional probability of back-back packet failure is higher due to temporal correlation. According to \cite{zorzi2}, the probability of back-back control packet failure is given as  
\begin{align}
    P_\text{bb} = \frac{1-\mathbb{E}_\gamma \lbrace P_e(1)\rbrace}{\mathbb{E}_\gamma \lbrace P_e(1)\rbrace} \left[ Q(\theta,\rho\theta) -Q(\rho\theta,\theta)] \right], 
\end{align}
where 
\begin{align}
\theta = \sqrt{ \frac{-2\log(1-\mathbb{E}_\gamma \lbrace P_e(1)\rbrace)}{1- \rho^2}}\\
\rho = \mathbf{J}_0(2\pi f_d T_s).
\end{align}
The probability that $n$ consecutive control packet fails is therefore given as 
\begin{align}
    P_e(n) =  P_e(n-1) \cdot P_\text{bb}
\end{align}
If $n_{max}$ is the maximum number of consecutive control packets failures that a GV can sustain, the system becomes unstable if more than $n_{max}$ consecutive packets fail \cite{AGVSCC}. 


Therefore, the probability of instability for coding rate $R$ is 
\begin{align}\label{eq:ch3.ieee.27}
    P_{us}(n_{max}, R) =  P_{e}(n_{max}(\nu,T_s)),
\end{align} 
where $P_e(n_{max}(\nu, T_s))$ is the probability of $n_{max}$ consecutive packet failures for a GV system with sampling time $T_s$, coding rate $R$ and with velocity $\nu$.

\begin{figure}
    \centering
    \input{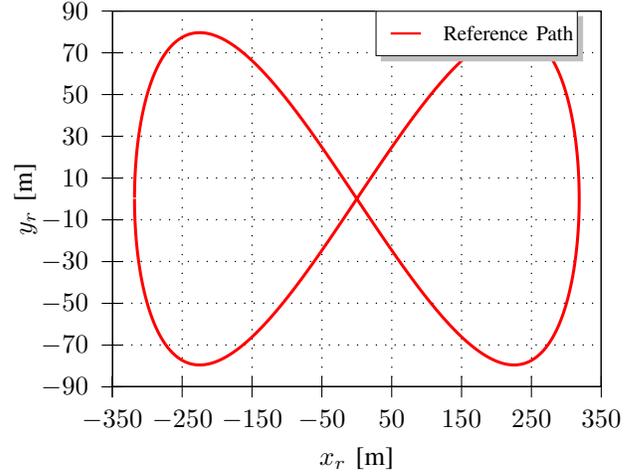}
    \caption{Reference Track of a GV}
    \label{fig:ch5:Results:16}
\end{figure}

\section{Simulation setup}
A system with $N$ GVs and channel bandwidth of 10 MHz is considered for the simulation. Every GV traces the path designed in Fig.~\ref{fig:ch5:Results:16}. An OFDMA system for downlink communication is considered and the resource elements are equally distributed among all the GVs. The resource element is the smallest time-frequency resource consisting of 1 subcarrier of 15KHz and 1 OFDM symbol.The total number of resource elements available for $i^{th}$ GV to transmit the control packet within time $T_s$ is given as,
   $ N_{rb,i} = \lfloor \frac{N_{RB} \times \lfloor \frac{T_s}{TTI}\rfloor} {N}  \rfloor$,
$N_{RB}$ is the total number of resource elements available per TTI. Each controller sends the control updates of $D_i = \unit[75]{bytes}$ with an interval of time $T_s$. A constant QPSK modulation scheme is used for packet transmission till the GV completes the track. 

\section{Results and discussion} \label{sec:ch5:results} 
\subsection{Probability of instability and coding rate}  \label{subsec:ch3:results}
\begin{figure}
    \centering
    \begingroup
\unitlength=1mm
\psset{xunit=65.00000mm, yunit=1.42857mm, linewidth=0.1mm}
\psset{arrowsize=2pt 3, arrowlength=1.4, arrowinset=.4}\psset{axesstyle=frame}
\begin{pspicture}(-0.23077, -46.20000)(1.00000, 0.00000)
\rput(-0.03077, -3.50000){%
\psaxes[subticks=0, labels=all, xsubticks=1, ysubticks=1, Ox=0, Oy=-35, Dx=0.2, Dy=10]{-}(0.00000, -35.00000)(0.00000, -35.00000)(1.00000, 0.00000)%
\multips(0.20000, -35.00000)(0.20000, 0.0){4}{\psline[linecolor=black, linestyle=dotted, linewidth=0.2mm](0, 0)(0, 35.00000)}
\multips(0.00000, -25.00000)(0, 10.00000){3}{\psline[linecolor=black, linestyle=dotted, linewidth=0.2mm](0, 0)(1.00000, 0)}
\rput[b](0.50000, -42.70000){Coding Rate \: $ R$ \: \unit{bpcu} }
\rput[t]{90}(-0.20000, -17.50000){$ \log_{10}(Pus) $}
\psclip{\psframe(0.00000, -35.00000)(1.00000, 0.00000)}
\psline[linecolor=blue, plotstyle=curve, linewidth=0.4mm, showpoints=true, linestyle=solid, linecolor=blue, dotstyle=o, dotscale=1.2 1.2, linewidth=0.4mm](0.03571, -41.02036)(0.07143, -32.51453)(0.10714, -27.74278)(0.14286, -24.49185)(0.17857, -22.06838)(0.21429, -20.16374)(0.25000, -18.61346)(0.28571, -17.31952)(0.32154, -16.21851)(0.35714, -15.26844)(0.39293, -14.43738)(0.42857, -13.70443)(0.46440, -13.05074)(0.50000, -12.46559)(0.53571, -11.93665)(0.57143, -11.45632)(0.60729, -11.01764)(0.64309, -10.61487)(0.67873, -10.24557)(0.71429, -9.90343)(0.75000, -9.58598)(0.78637, -9.29024)(0.82192, -9.01447)(0.85714, -8.75661)(0.89286, -8.51433)(0.92879, -8.28503)(0.96463, -8.07127)(1.00000, -7.86838)
\psline[linecolor=red, plotstyle=curve, linewidth=0.4mm, showpoints=true, linestyle=solid, linecolor=red, dotstyle=diamond, dotscale=1.2 1.2, linewidth=0.4mm](0.03571, -56.97560)(0.07143, -48.04369)(0.10714, -42.85975)(0.14286, -39.21016)(0.17857, -36.40123)(0.21429, -34.12393)(0.25000, -32.21344)(0.28571, -30.57136)(0.32154, -29.13379)(0.35714, -27.85869)(0.39293, -26.71327)(0.42857, -25.67687)(0.46440, -24.72943)(0.50000, -23.86102)(0.53571, -23.05799)(0.57143, -22.31271)(0.60729, -21.61774)(0.64309, -20.96683)(0.67873, -20.35859)(0.71429, -19.78484)(0.75000, -19.24326)(0.78637, -18.73043)(0.82192, -18.24475)(0.85714, -17.78392)(0.89286, -17.34489)(0.92879, -16.92386)(0.96463, -16.52647)(1.00000, -16.14485)
\psline[linecolor=darkred, plotstyle=curve, linewidth=0.4mm, showpoints=true, linestyle=solid, linecolor=darkred, dotstyle=square, dotscale=1.2 1.2, linewidth=0.4mm](0.03571, -65.33415)(0.07143, -56.32112)(0.10714, -51.05644)(0.14286, -47.32650)(0.17857, -44.43760)(0.21429, -42.08074)(0.25000, -40.09109)(0.28571, -38.37026)(0.32154, -36.85432)(0.35714, -35.50132)(0.39293, -34.27837)(0.42857, -33.16494)(0.46440, -32.14079)(0.50000, -31.19626)(0.53571, -30.31748)(0.57143, -29.49691)(0.60729, -28.72704)(0.64309, -28.00159)(0.67873, -27.31960)(0.71429, -26.67241)(0.75000, -26.05784)(0.78637, -25.47241)(0.82192, -24.91469)(0.85714, -24.38238)(0.89286, -23.87227)(0.92879, -23.38021)(0.96463, -22.91309)(1.00000, -22.46193)
\psline[linecolor=cyan, plotstyle=curve, linewidth=0.4mm, showpoints=true, linestyle=solid, linecolor=cyan, dotstyle=triangle, dotscale=1.2 1.2, linewidth=0.4mm](0.03571, -70.07950)(0.07143, -61.03943)(0.10714, -55.74768)(0.14286, -51.99064)(0.17857, -49.07461)(0.21429, -46.69060)(0.25000, -44.67376)(0.28571, -42.92572)(0.32154, -41.38254)(0.35714, -40.00230)(0.39293, -38.75206)(0.42857, -37.61134)(0.46440, -36.55985)(0.50000, -35.58803)(0.53571, -34.68190)(0.57143, -33.83396)(0.60729, -33.03671)(0.64309, -32.28382)(0.67873, -31.57450)(0.71429, -30.89991)(0.75000, -30.25793)(0.78637, -29.64505)(0.82192, -29.05991)(0.85714, -28.50021)(0.89286, -27.96267)(0.92879, -27.44303)(0.96463, -26.94863)(1.00000, -26.47007)
\psline[linecolor=black, plotstyle=curve, linewidth=0.4mm, showpoints=true, linestyle=solid, linecolor=black, dotstyle=o, dotscale=1.2 1.2, linewidth=0.4mm](0.03571, -72.54557)(0.07143, -63.49499)(0.10714, -58.19270)(0.14286, -54.42509)(0.17857, -51.49845)(0.21429, -49.10380)(0.25000, -47.07629)(0.28571, -45.31754)(0.32154, -43.76362)(0.35714, -42.37262)(0.39293, -41.11157)(0.42857, -39.96003)(0.46440, -38.89766)(0.50000, -37.91496)(0.53571, -36.99791)(0.57143, -36.13902)(0.60729, -35.33077)(0.64309, -34.56685)(0.67873, -33.84651)(0.71429, -33.16085)(0.75000, -32.50777)(0.78637, -31.88376)(0.82192, -31.28746)(0.85714, -30.71659)(0.89286, -30.16784)(0.92879, -29.63689)(0.96463, -29.13128)(1.00000, -28.64145)
\endpsclip
\psframe[linecolor=black, fillstyle=solid, fillcolor=white, shadowcolor=lightgray, shadowsize=1mm, shadow=true](0.07692, -14.00000)(0.46154, 0.00000)
\rput[l](0.21538, -2.10000){\footnotesize{$\text{v = 1m/s}$}}
\psline[linecolor=blue, linestyle=solid, linewidth=0.3mm](0.10769, -2.10000)(0.16923, -2.10000)
\psline[linecolor=blue, linestyle=solid, linewidth=0.3mm](0.10769, -2.10000)(0.16923, -2.10000)
\psdots[linecolor=blue, linestyle=solid, linewidth=0.3mm, dotstyle=o, dotscale=1.2 1.2, linecolor=blue](0.13846, -2.10000)
\rput[l](0.21538, -4.55000){\footnotesize{$\text{v = 2m/s}$}}
\psline[linecolor=red, linestyle=solid, linewidth=0.3mm](0.10769, -4.55000)(0.16923, -4.55000)
\psline[linecolor=red, linestyle=solid, linewidth=0.3mm](0.10769, -4.55000)(0.16923, -4.55000)
\psdots[linecolor=red, linestyle=solid, linewidth=0.3mm, dotstyle=diamond, dotscale=1.2 1.2, linecolor=red](0.13846, -4.55000)
\rput[l](0.21538, -7.00000){\footnotesize{$\text{v = 3m/s}$}}
\psline[linecolor=darkred, linestyle=solid, linewidth=0.3mm](0.10769, -7.00000)(0.16923, -7.00000)
\psline[linecolor=darkred, linestyle=solid, linewidth=0.3mm](0.10769, -7.00000)(0.16923, -7.00000)
\psdots[linecolor=darkred, linestyle=solid, linewidth=0.3mm, dotstyle=square, dotscale=1.2 1.2, linecolor=darkred](0.13846, -7.00000)
\rput[l](0.21538, -9.45000){\footnotesize{$\text{v = 4m/s}$}}
\psline[linecolor=cyan, linestyle=solid, linewidth=0.3mm](0.10769, -9.45000)(0.16923, -9.45000)
\psline[linecolor=cyan, linestyle=solid, linewidth=0.3mm](0.10769, -9.45000)(0.16923, -9.45000)
\psdots[linecolor=cyan, linestyle=solid, linewidth=0.3mm, dotstyle=triangle, dotscale=1.2 1.2, linecolor=cyan](0.13846, -9.45000)
\rput[l](0.21538, -11.90000){\footnotesize{$\text{v = 5m/s}$}}
\psline[linecolor=black, linestyle=solid, linewidth=0.3mm](0.10769, -11.90000)(0.16923, -11.90000)
\psline[linecolor=black, linestyle=solid, linewidth=0.3mm](0.10769, -11.90000)(0.16923, -11.90000)
\psdots[linecolor=black, linestyle=solid, linewidth=0.3mm, dotstyle=o, dotscale=1.2 1.2, linecolor=black](0.13846, -11.90000)
}
\end{pspicture}
\endgroup
 
    \caption{Probability of instability, $T_s = \unit[1]{ms}$}
    \label{fig:ch5:Results:9}
\end{figure}
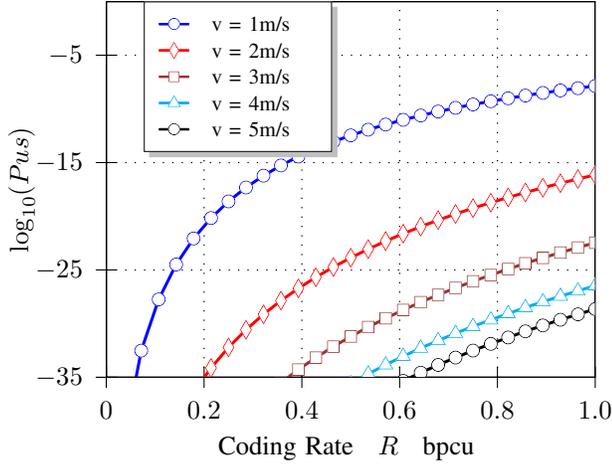

The likelihood of a WNCS being unstable, when controlled by a centralized controller over a Rayleigh fading channel is the probability of instability $P_{us}$. The probability that a WNCS is stable depends on the sampling time, GV velocity and the coding rate. In this section, we discuss the impact of coding rate on the stability performance of a WNCS. 
Fig.~\ref{fig:ch5:Results:9} shows the behavior of $P_{us}$ over coding rate for different GV velocities $\unit[1]{m/s}-\unit[5]{m/s}$. It shows that $P_{us}$ decreases at lower coding rate. At lower coding rate the number of redundant bits is high and the packet is less prone to errors caused by channel fading. Therefore, the block error probability decreases at lower coding rate, hence further reducing $P_{us}$. In order to have a highly stable wireless control system of a GV, a packet should be transmitted with a lower coding rate. However, encoding a packet with lower coding rate will consume more communication resources. Additionally, along with the coding rate, $P_{us}$ also depends upon GV's velocity and the sampling time. As shown in Fig.~\ref{fig:ch5:Results:9},  $P_{us}$ decreases as the GV velocity increases from \unit[1]{m/s} to \unit[5]{m/s}. At lower velocity the channel gain over adjacent packets is highly correlated, particularly for short packet transmission. A WNCS is more likely to become unstable if large number of consecutive packets are lost ~\eqref{eq:ch3.ieee.27}. At higher velocity, the channel correlation decreases, hence, the probability of consecutive packet loss is lower. It illustrates that to reduce the probability of instability without increasing the resource consumption, the GV should drive at higher speed. Higher the velocity, lower is the channel correlation and hence, lower is the $P_{us}$. 


\paragraph*{Admissible coding rate}
$P_{us}$ determines the reliability and robustness of a WNCS. To assure a highly stable WNCS, $P_{us}$ should be small and less than the threshold probability $P_{us}^{th}$ defined according to the industrial application requirements. We evaluate the admissible coding rate $R_{req}$, required to have a stable wireless control system so that $P_{us}$ is less than $P_{us}^{th}$. The coding rate required by GV control system with sampling time $T_s$ and velocity $\nu$ is  
\begin{equation}
  R_{req} = \lbrace{ \max R; \: \text{s.t.} \: P_{us}(n_{max},R) \leq P_{us}^{th} \rbrace}.
\end{equation}

\begin{figure}
    \centering
      \begingroup
\unitlength=1mm
\psset{xunit=14.44444mm, yunit=45.45455mm, linewidth=0.1mm}
\psset{arrowsize=2pt 3, arrowlength=1.4, arrowinset=.4}\psset{axesstyle=frame}
\begin{pspicture}(-0.03846, -0.35200)(5.50000, 1.10000)
\rput(-0.13846, -0.11000){%
\psaxes[subticks=0, labels=all, xsubticks=1, ysubticks=1, Ox=1, Oy=0, Dx=1, Dy=0.2]{-}(1.00000, 0.00000)(1.00000, 0.00000)(5.50000, 1.10000)%
\multips(2.00000, 0.00000)(1.00000, 0.0){4}{\psline[linecolor=black, linestyle=dotted, linewidth=0.2mm](0, 0)(0, 1.10000)}
\multips(1.00000, 0.20000)(0, 0.20000){5}{\psline[linecolor=black, linestyle=dotted, linewidth=0.2mm](0, 0)(4.50000, 0)}
\rput[b](3.25000, -0.24200){$\text{Velocity} \: \nu \: [\unit{m/s}] $}
\rput[t]{90}(0.10000, 0.55000){$ R_{req} [\unit{bpcu}] $}
\psclip{\psframe(1.00000, 0.00000)(5.50000, 1.10000)}
\psline[linecolor=blue, plotstyle=curve, linewidth=0.4mm, showpoints=true, linestyle=solid, linecolor=blue, dotstyle=o, dotscale=1.2 1.2, linewidth=0.4mm](1.00000, 0.13953)(2.00000, 0.40000)(3.00000, 0.66667)(4.00000, 1.00000)(5.00000, 1.00000)
\psline[linecolor=red, plotstyle=curve, linewidth=0.4mm, showpoints=true, linestyle=solid, linecolor=red, dotstyle=diamond, dotscale=1.2 1.2, linewidth=0.4mm](1.00000, 0.12245)(2.00000, 0.14634)(3.00000, 0.28571)(4.00000, 0.42857)(5.00000, 0.54545)
\psline[linecolor=darkred, plotstyle=curve, linewidth=0.4mm, showpoints=true, linestyle=solid, linecolor=darkred, dotstyle=square, dotscale=1.2 1.2, linewidth=0.4mm](1.00000, 0.12245)(2.00000, 0.12245)(3.00000, 0.16667)(4.00000, 0.26087)(5.00000, 0.35294)
\psline[linecolor=cyan, plotstyle=curve, linewidth=0.4mm, showpoints=true, linestyle=solid, linecolor=cyan, dotstyle=triangle, dotscale=1.2 1.2, linewidth=0.4mm](1.00000, 0.12245)(2.00000, 0.12245)(3.00000, 0.12245)(4.00000, 0.16667)(5.00000, 0.24000)
\psline[linecolor=black, plotstyle=curve, linewidth=0.4mm, showpoints=true, linestyle=solid, linecolor=black, dotstyle=o, dotscale=1.2 1.2, linewidth=0.4mm](1.00000, 0.12245)(2.00000, 0.12245)(3.00000, 0.12245)(4.00000, 0.13953)(5.00000, 0.19355)
\endpsclip
\psframe[linecolor=black, fillstyle=solid, fillcolor=white, shadowcolor=lightgray, shadowsize=1mm, shadow=true](1.34615, 0.66000)(3.07692, 1.10000)
\rput[l](1.96923, 1.03400){\footnotesize{$T_s = \unit[1]{ms}$}}
\psline[linecolor=blue, linestyle=solid, linewidth=0.3mm](1.48462, 1.03400)(1.76154, 1.03400)
\psline[linecolor=blue, linestyle=solid, linewidth=0.3mm](1.48462, 1.03400)(1.76154, 1.03400)
\psdots[linecolor=blue, linestyle=solid, linewidth=0.3mm, dotstyle=o, dotscale=1.2 1.2, linecolor=blue](1.62308, 1.03400)
\rput[l](1.96923, 0.95700){\footnotesize{$T_s = \unit[2]{ms}$}}
\psline[linecolor=red, linestyle=solid, linewidth=0.3mm](1.48462, 0.95700)(1.76154, 0.95700)
\psline[linecolor=red, linestyle=solid, linewidth=0.3mm](1.48462, 0.95700)(1.76154, 0.95700)
\psdots[linecolor=red, linestyle=solid, linewidth=0.3mm, dotstyle=diamond, dotscale=1.2 1.2, linecolor=red](1.62308, 0.95700)
\rput[l](1.96923, 0.88000){\footnotesize{$T_s = \unit[3]{ms}$}}
\psline[linecolor=darkred, linestyle=solid, linewidth=0.3mm](1.48462, 0.88000)(1.76154, 0.88000)
\psline[linecolor=darkred, linestyle=solid, linewidth=0.3mm](1.48462, 0.88000)(1.76154, 0.88000)
\psdots[linecolor=darkred, linestyle=solid, linewidth=0.3mm, dotstyle=square, dotscale=1.2 1.2, linecolor=darkred](1.62308, 0.88000)
\rput[l](1.96923, 0.80300){\footnotesize{$T_s = \unit[4]{ms}$}}
\psline[linecolor=cyan, linestyle=solid, linewidth=0.3mm](1.48462, 0.80300)(1.76154, 0.80300)
\psline[linecolor=cyan, linestyle=solid, linewidth=0.3mm](1.48462, 0.80300)(1.76154, 0.80300)
\psdots[linecolor=cyan, linestyle=solid, linewidth=0.3mm, dotstyle=triangle, dotscale=1.2 1.2, linecolor=cyan](1.62308, 0.80300)
\rput[l](1.96923, 0.72600){\footnotesize{$T_s = \unit[5]{ms}$}}
\psline[linecolor=black, linestyle=solid, linewidth=0.3mm](1.48462, 0.72600)(1.76154, 0.72600)
\psline[linecolor=black, linestyle=solid, linewidth=0.3mm](1.48462, 0.72600)(1.76154, 0.72600)
\psdots[linecolor=black, linestyle=solid, linewidth=0.3mm, dotstyle=o, dotscale=1.2 1.2, linecolor=black](1.62308, 0.72600)
}\end{pspicture}
\endgroup
 
    \caption{Admissible coding rate for $P_{us}^{th}$ = $10^{-9}$}
    \label{fig:ch5:Results:11}
\end{figure}
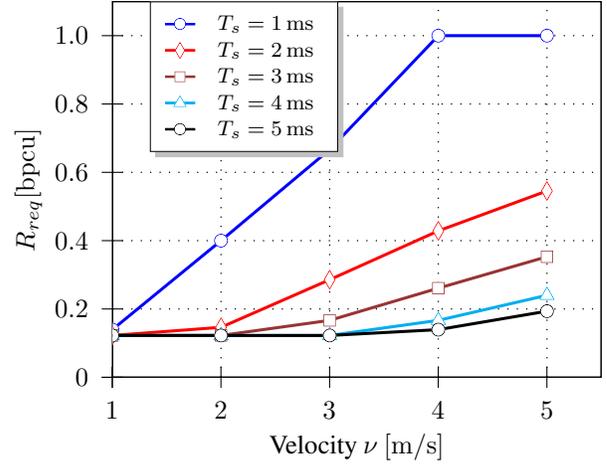

Fig.~\ref{fig:ch5:Results:11} shows the admissible coding rate to assure a stable wireless control system. The admissible coding rate to achieve the threshold $P_{us}^{th}$, increases as the GV velocity is increased from $\unit[1]{m/s}$ to $\unit[5]{m/s}$. Moreover, Fig.~\ref{fig:ch5:Results:11} illustrates that the admissible coding rate increases if $T_s$ is reduced from $\unit[5]{ms}$ to $\unit[1]{ms}$. At higher velocity, $P_{us}$ decreases due to lower channel correlation between the adjacent control packets. Therefore, at higher GV velocity, encoding the packet with a higher coding rate,  gives the same stability performance of a WNCS.

  Furthermore, lower sampling time implies frequent transmission of the control updates in the downlink i.\,e., lower variation between the consecutive updates. A GV control system can sustain higher number of consecutive packet drops i.\,e. higher outage tolerance at lower sampling time. Additionally, the probability that such a large number of consecutive packets drop is lower. Hence, $P_{us}$ is low at lower sampling time and thus the packet can be encoded with higher coding rate. In Fig.~\ref{fig:ch5:Results:11}, at $T_s = \unit[1]{ms}$ and $\nu > \unit[4]{m/s}$, the coding rate can be as high as \unit[1]{bpcu} to maintain the WNCS instability below $10^{-9}$. On the contrary, as the sampling time increases, the variation between two consecutive control updates also increases. Hence, the outage tolerance reduces at higher sampling time and higher is the $P_{us}$. Therefore, in order to maintain a stable wireless control at a higher sampling time, a packet must be sent with high reliability i.e. with lower coding rate.
  
\subsection{Maximum number of supported GVs}
In this section, we evaluate the maximum number of GV systems, $N_{max}$, that can be supported with the available communication resources, assuring that the probability of instability is within the threshold. 
The maximum number of GVs supported is evaluated as
\begin{align}
  N_{max} =  \lbrace  \max N; \: \text{s.t.} \: P_{us}\left(n_{max}, R\right) < P_{us}^{th} \rbrace.
\end{align} 

\begin{figure}
    \centering
    \begingroup
\unitlength=1mm
\psset{xunit=10.83333mm, yunit=0.23810mm, linewidth=0.1mm}
\psset{arrowsize=2pt 3, arrowlength=1.4, arrowinset=.4}\psset{axesstyle=frame}
\begin{pspicture}(-13.38462, -67.20000)(-6.00000, 210.00000)
\rput(-0.18462, -21.00000){%
\psaxes[subticks=0, labels=all, xsubticks=1, ysubticks=1, Ox=-12, Oy=0, Dx=2, Dy=50]{-}(-12.00000, 0.00000)(-12.00000, 0.00000)(-6.00000, 210.00000)%
\multips(-10.00000, 0.00000)(2.00000, 0.0){2}{\psline[linecolor=black, linestyle=dotted, linewidth=0.2mm](0, 0)(0, 210.00000)}
\multips(-12.00000, 50.00000)(0, 50.00000){3}{\psline[linecolor=black, linestyle=dotted, linewidth=0.2mm](0, 0)(6.00000, 0)}
\rput[b](-9.00000, -46.20000){$ \log_{10}(P_{us}^{th}) $}
\rput[t]{90}(-13.20000, 105.00000){$ N_{max} $}
\psline[linecolor=blue, plotstyle=curve, linewidth=0.4mm, showpoints=true, linestyle=solid, linecolor=blue, dotstyle=o, dotscale=1.2 1.2, linewidth=0.4mm](-12.00000, 20.00000)(-11.00000, 21.00000)(-10.00000, 23.00000)(-9.00000, 25.00000)(-8.00000, 27.00000)(-7.00000, 28.00000)(-6.00000, 28.00000)
\psline[linecolor=red, plotstyle=curve, linewidth=0.4mm, showpoints=true, linestyle=solid, linecolor=red, dotstyle=diamond, dotscale=1.2 1.2, linewidth=0.4mm](-12.00000, 41.00000)(-11.00000, 44.00000)(-10.00000, 47.00000)(-9.00000, 50.00000)(-8.00000, 54.00000)(-7.00000, 56.00000)(-6.00000, 56.00000)
\psline[linecolor=darkred, plotstyle=curve, linewidth=0.4mm, showpoints=true, linestyle=solid, linecolor=darkred, dotstyle=square, dotscale=1.2 1.2, linewidth=0.4mm](-12.00000, 63.00000)(-11.00000, 67.00000)(-10.00000, 72.00000)(-9.00000, 76.00000)(-8.00000, 82.00000)(-7.00000, 84.00000)(-6.00000, 84.00000)
\psline[linecolor=cyan, plotstyle=curve, linewidth=0.4mm, showpoints=true, linestyle=solid, linecolor=cyan, dotstyle=triangle, dotscale=1.2 1.2, linewidth=0.4mm](-12.00000, 77.00000)(-11.00000, 83.00000)(-10.00000, 89.00000)(-9.00000, 96.00000)(-8.00000, 103.00000)(-7.00000, 112.00000)(-6.00000, 112.00000)
\psline[linecolor=black, plotstyle=curve, linewidth=0.4mm, showpoints=true, linestyle=solid, linecolor=black, dotstyle=o, dotscale=1.2 1.2, linewidth=0.4mm](-12.00000, 94.00000)(-11.00000, 102.00000)(-10.00000, 109.00000)(-9.00000, 118.00000)(-8.00000, 128.00000)(-7.00000, 138.00000)(-6.00000, 140.00000)
\psframe[linecolor=black, fillstyle=solid, fillcolor=white, shadowcolor=lightgray, shadowsize=1mm, shadow=true](-11.69231, 126.00000)(-9.46154, 210.00000)
\rput[l](-10.86154, 197.40000){\footnotesize{$T_s = \unit[1]{ms}$}}
\psline[linecolor=blue, linestyle=solid, linewidth=0.3mm](-11.50769, 197.40000)(-11.13846, 197.40000)
\psline[linecolor=blue, linestyle=solid, linewidth=0.3mm](-11.50769, 197.40000)(-11.13846, 197.40000)
\psdots[linecolor=blue, linestyle=solid, linewidth=0.3mm, dotstyle=o, dotscale=1.2 1.2, linecolor=blue](-11.32308, 197.40000)
\rput[l](-10.86154, 182.70000){\footnotesize{$T_s = \unit[2]{ms}$}}
\psline[linecolor=red, linestyle=solid, linewidth=0.3mm](-11.50769, 182.70000)(-11.13846, 182.70000)
\psline[linecolor=red, linestyle=solid, linewidth=0.3mm](-11.50769, 182.70000)(-11.13846, 182.70000)
\psdots[linecolor=red, linestyle=solid, linewidth=0.3mm, dotstyle=diamond, dotscale=1.2 1.2, linecolor=red](-11.32308, 182.70000)
\rput[l](-10.86154, 168.00000){\footnotesize{$T_s = \unit[3]{ms}$}}
\psline[linecolor=darkred, linestyle=solid, linewidth=0.3mm](-11.50769, 168.00000)(-11.13846, 168.00000)
\psline[linecolor=darkred, linestyle=solid, linewidth=0.3mm](-11.50769, 168.00000)(-11.13846, 168.00000)
\psdots[linecolor=darkred, linestyle=solid, linewidth=0.3mm, dotstyle=square, dotscale=1.2 1.2, linecolor=darkred](-11.32308, 168.00000)
\rput[l](-10.86154, 153.30000){\footnotesize{$T_s = \unit[4]{ms}$}}
\psline[linecolor=cyan, linestyle=solid, linewidth=0.3mm](-11.50769, 153.30000)(-11.13846, 153.30000)
\psline[linecolor=cyan, linestyle=solid, linewidth=0.3mm](-11.50769, 153.30000)(-11.13846, 153.30000)
\psdots[linecolor=cyan, linestyle=solid, linewidth=0.3mm, dotstyle=triangle, dotscale=1.2 1.2, linecolor=cyan](-11.32308, 153.30000)
\rput[l](-10.86154, 138.60000){\footnotesize{$T_s = \unit[5]{ms}$}}
\psline[linecolor=black, linestyle=solid, linewidth=0.3mm](-11.50769, 138.60000)(-11.13846, 138.60000)
\psline[linecolor=black, linestyle=solid, linewidth=0.3mm](-11.50769, 138.60000)(-11.13846, 138.60000)
\psdots[linecolor=black, linestyle=solid, linewidth=0.3mm, dotstyle=o, dotscale=1.2 1.2, linecolor=black](-11.32308, 138.60000)
}\end{pspicture}
\endgroup
 
    \caption{Maximum supported GV, $\nu$ = \unit[5]{m/s}}
    \label{fig:ch5:Results:15}
\end{figure}
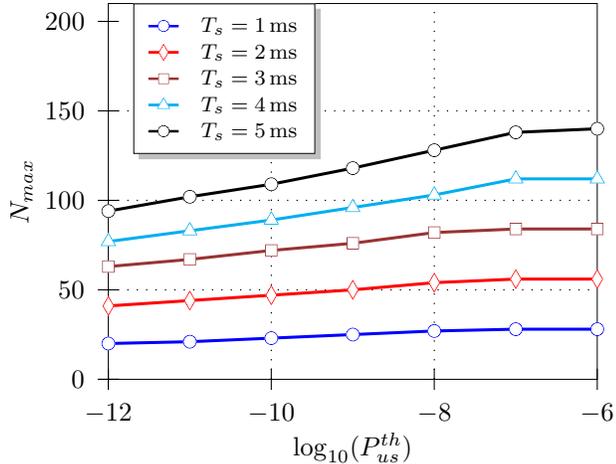

$P_{us}$ determines the instability and robustness of a wireless controlled system. Fig.~\ref{fig:ch5:Results:15} shows the maximum number of GVs that can be scheduled, while simultaneously satisfying the WNCS stability constraints. If the control system stability requirement is lower, i.e. at higher $P_{us}^{th}$,  higher number of GVs are supported. The number of GV supported decreases if the stability constraints on the control system are becoming more strict with lower $P_{us}^{th}$. The reason for this behavior is that with constant sampling time and GV velocity, the probability of instability can be reduced by lowering the coding rate. Lowering the coding rate would need more resource elements to transmit the same number of information bits. As the available resources are not sufficient to satisfy the stability constraint for all the GVs, lower number of GVs are admitted. Therefore, the number of supported GVs decreases as $P_{us}^{th}$ is reduced from $10^{-6}$ to $10^{-12}$.

Fig.~\ref{fig:ch5:Results:15} shows that the $N_{max}$ increases from $25$ GVs to $95$ GVs if $T_s$ is increased from $\unit[1]{ms}$ to $\unit[5]{ms}$. A control packet must be transmitted and received at the actuator within the control system sampling time. The decoded packet is outdated if the latency is higher than the control system sampling time. If the sampling time is increased, the time within which the control update should reach the GV increases. This allows the cloud controller to allocate more resources to a GV and schedule the transmission of remaining GVs at the next TTI. Therefore, the maximum number of GVs that can be supported increases at higher sampling time, while still retaining the instability probability within the threshold. 


\section{Conclusion} \label{sec:ch5:conclusion}
We discussed a practically relevant use-case of a centralized cloud controlling multiple GVs, to study the impact of the communication and control system parameters on the stability performance. Our work investigates the impact of the short block codes on the stability of a non-linear time varying GV control system. It is evident that the stability of a WNCS strongly depends on the coding rate and the packet error probability of a wireless link. However, our findings suggest that communication resource consumption can be reduced by adapting the control parameters like sampling time and the GV velocity while maintaining a constant stability performance. Moreover, the paper also presents an analysis on the maximum number of admissible GVs with respect to the stability and latency constraints of the WNCS. 
Our research work focuses on the impact of downlink communication on the stability of a GV control system. In future, the framework can be extended to study the impact of uplink feedback channel and analyse the conditional dependencies of a uplink-downlink communication on the stability performance. Moreover, the results of this paper have indicated the existing potential to increase the admissible number of GVs by optimally allocating the resources. We plan to further investigate the resource allocation strategies considering both the control and communication aspects. 

\section{Acknowledgment}
This research was supported by the German Federal Ministry of Education and Research (BMBF) under grant number
KIS15GTI007 (project TACNET4.0, www.tacnet40.de). The responsibility for this publication lies with the authors.


\balance
\bibliographystyle{abbrv}
\bibliography{bibiliography}
\end{document}